\documentclass[10pt,letterpaper]{article}
\usepackage{opex3}
\usepackage{amsmath,amssymb}
\usepackage{graphicx}

\newcommand{\ve}{\varepsilon}
\newcommand{\veh}{\varepsilon_h}

\newcommand{\om}{\omega}
\newcommand{\prp}{\scriptscriptstyle{\perp}}
\newcommand{\vrt}{\scriptscriptstyle{\Vert}}

\begin{document}


\title{Subwavelength plasmonic kinks in arrays of metallic nanoparticles}

\author{Roman E. Noskov,$^{1,*}$ Pavel A. Belov,$^1$ and Yuri S. Kivshar$^{1,2}$}

\address{$^{1}$National Research University of Information Technologies, Mechanics and Optics, St. Petersburg 197101, Russia \\
$^{2}$Nonlinear Physics Centre, Research School of Physics and Engineering, Australian National University, Canberra ACT 0200, Australia}

\email{$^*$nanometa@gmail.com} 

\begin{abstract}
We analyze nonlinear effects in optically driven arrays of nonlinear metallic nanoparticles. We demonstrate
that such plasmonic systems are characterized by a bistable response, and they can support the propagation
of dissipative switching waves (or plasmonic kinks) connecting the states with different polarization.
We study numerically the properties of such plasmonic kinks which are characterized by a subwavelength extent
and a tunable velocity.
\end{abstract}




\section{Introduction}

When the size of conventional optical devices is reduced to nanoscale,
the spatial confinement of light becomes inherently limited by diffraction.
However, metal-dielectric structures allow to achieve a subwavelength light confinement
with surface plasmon polaritons (SPPs), or {\em plasmons}~\cite{r1a,r1b,r1c}. In particular, by using
plasmons excited in an array of resonantly coupled metal nanoparticles~\cite{r2a,r2b},
one can spatially confine and manipulate optical energy over distances much smaller than the wavelength.

Because of the strong enhancement of the field induced by the excitation of SPPs and
increased optical nonlinearity, surface plasmons can be employed for the realization
of a variety of nonlinear optical effects. In particular, several nonlinear optical processes
have been demonstrated in plasmonic nanostructures, e.g., optical limiting and self-phase modulation
in arrays of structured nanoparticles~\cite{r3}, second-harmonic generation and bistability in nanostructured metal films~\cite{r4,r5,zayats_prl} and terahertz radiation generation in nonlinear metallic nanodimers~\cite{noskov_jap,noskov_prb}.
In addition, strong geometric confinement can boost efficiency of nonlinear optical effects,
including the existence of subwavelength solitons in  metal-dielectric multilayers~\cite{prl_zhang,skryabin_oe_2011,skryabin_ol_2010}
and arrays of metal nanowires~\cite{prl_panoiu}. The solitons supported by such media result from
a balance between tunneling of surface plasmon modes and nonlinear self-trapping.

Here we study nonlinear effects in optically-driven arrays of nonlinear metal nanoparticles. In addition to the subwavelength plasmonic solitons analyzed earlier in other plasmonic systems~\cite{prl_zhang,skryabin_oe_2011,skryabin_ol_2010,prl_panoiu}, we demonstrate that an array of nonlinear metal
nanoparticles can support the propagation of dissipative switching waves (or plasmonic kinks) connecting the states with different polarization.
These kinks become possible due to a bistable nonlinear response of each nanoparticle \cite{leung_pra}, and they have a subwavelength extent
and a finite (or zero) velocity.

The paper is organized as follows. In Sec.~2, we describe our nonlinear model and discuss its realistic parameters. Section~3 summarizes our results
for the study of the basic properties of plasmonic kinks and their generation. Finally, Sec.~4 concludes the paper.

\section{Model and basic equations}

We consider a linear array of identical spherical metallic (silver) nanoparticles embedded into a SiO$_2$ host medium with permittivity $\veh$, and is exited by the optical field ${\bf E}^{(ex)}$ (see Fig.~1). We assume that the particle radii and the center-to-center distance are $a=10$~nm and $d=30$~nm, respectively. Ratio $a/d$ satisfies the condition $a/d \leq1/3$, so that we can employ the point dipole approximation~\cite{Yong}. In the optical spectral range, a linear part of silver dielectric constant can be written in a generalized Drude form $\ve_{\text{Ag}}^\text{L}=\ve_{\infty}-\om_p^2/[\om(\om-i\nu)]$, where $\ve_\infty=4.96$, $\hbar\om_p=9.54$~eV, $\hbar\nu=0.055$~eV~\cite{Johnson} (hereinafter we accept $\exp(i\om t)$ time dependence); whereas dispersion of SiO$_2$ can be neglected since $\veh\simeq2.15$ for wavelengths 350 –- 450~nm~\cite{Palik}. Nonlinear dielectric constant of silver is $\ve_{\text{Ag}}^{\text{NL}}=\ve_{\text{Ag}}^\text{L}+\chi^{(3)}|{\bf E}^{(in)}_n|^2$, where ${\bf E}^{(in)}_n$ is the local field inside $n$-th particle. We keep only cubic susceptibility due to spherical symmetry of particles. According to the model suggested in Ref.~\cite{Drachev} and confirmed in experiment, 10 nm radii Ag spheres possess a remarkably high and purely real cubic susceptibility $\chi^{(3)}\simeq 3\times 10^{-9}$~esu, in comparing to which the cubic nonlinearity of SiO$_2$ is negligibly weak ($\thicksim10^{-15}$~esu~\cite{weber_book_03}).

\begin{figure}
\centerline{\mbox{\resizebox{7.5cm}{!}{\includegraphics{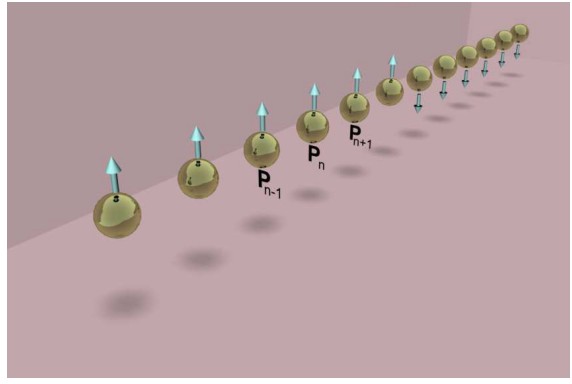}}}}
\caption{\label{fig:1} Schematic sketch of
an array of coupled metallic particles with a nonlinear response.
Arrows indicate a profile of the polarizations for a typical kink configuration.}\label{fig1}
\end{figure}

To study nonlinear dynamics of the particle polarizations, ${\bf p}_{n}$, we employ the model based on the dispersion relation method. This technique was firstly applied to investigation of light scattering by nonlinear metal nanodimers \cite{noskov_jap,noskov_prb} and nonlinear effects in magnetic metamaterials \cite{Zharov,Rosanov}, and here we extend it to the case of an array of metal nanoparticle. According to this model, temporal dynamical response of metal nanoparticle arrays excited by the optical field with the frequency lying close to the frequency of the surface plasmon resonance of an individual particle, $\om_0=\om_p/\sqrt{\ve_\infty + 2\veh}$, can be described by the following system of coupled equations for
slowly varying amplitudes of the particle dipole moments:
\begin{equation}\label{dynamic}
\begin{split}
-i\frac{d P_n^{\prp}}{d\tau}+\left(-i\gamma+\Omega+|{\bf P}_n|^2 \right) P_n^{\prp}+ \sum_{m\neq n} G_{n,m}^{\prp} P_m^{\prp} &= E_n^{\prp}, \\
-i\frac{d P_n^{\vrt}}{d\tau}+\left(-i\gamma+\Omega+|{\bf P}_n|^2 \right) P_n^{\vrt}+ \sum_{m\neq n} G_{n,m}^{\vrt} P_m^{\vrt} &= E_n^{\vrt},
\end{split}
\end{equation}
where
$$
G_{n,m}^{\prp}  = \frac{\eta}{2} \left( (k_0 d)^2 - \frac{i k_0 d}{|n-m|}- \frac{1}{|n-m|^2} \right) \frac{\exp(-i k_0 d|n-m|)}{|n-m|}$$ and
$$
G_{n,m}^{\vrt}  = \eta \left(\frac{i k_0 d}{|n-m|} + \frac{1}{|n-m|^2} \right) \frac{\exp(-i k_0 d|n-m|)}{|n-m|},
$$
are in charge of dipole-dipole interactions between particles, $P_n^{\prp,\vrt}=p_n^{\prp,\vrt}\sqrt{\chi^{(3)}}/(\sqrt{2(\ve_\infty+2\veh)}\veh a^3)$ and $E_n^{\prp,\vrt} = -3 \veh \sqrt{\chi^{(3)}} E^{(ex)\prp,\vrt}_n/\sqrt{8(\ve_\infty+2\veh)^3}$ are dimensionless slowly varying amplitudes of the particle dipole moments and external electric field, respectively, the indices '$\prp$' and '$\vrt$' stand for the transverse and longitudinal components with respect to the array direction,
$\eta=\frac{3\veh}{\ve_{\infty}+2\veh}\left(\frac{a}{d}\right)^3$, $|{\bf P}_n|^2=|P_n^{\prp}|^2+|P_n^{\vrt}|^2$,  $\gamma=\nu/(2\om_0)+(k_0 a)^3\veh/(\ve_\infty+2\veh)$ describes both thermal and radiation losses of particles, $k_0=\om_0/c\sqrt{\veh}$, $\Omega=(\om-\om_0)/\om_0$ and $\tau = \om_0 t$. Equations~(\ref{dynamic}) describe
temporal nonlinear dynamics of an array of metal nanoparticles driven by arbitrary external optical field with the frequency $\om\sim\om_0$.
We stress that model takes into account all particle interactions through the dipole fields, and it can be applied both to finite and
infinite arrays.

\section{Bistable response and plasmonic kinks}

\begin{figure}
\centerline{\mbox{\resizebox{13cm}{!}{\includegraphics{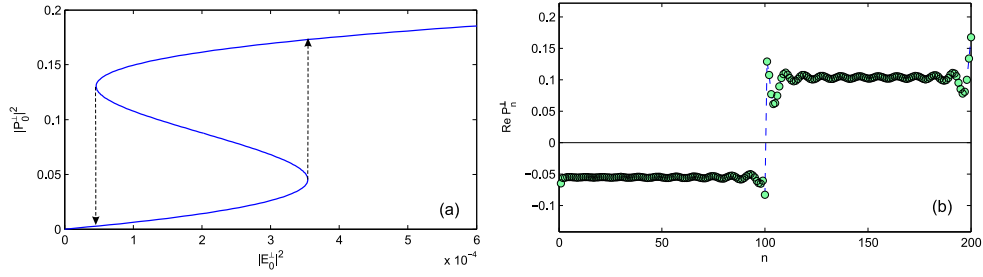}}}}
\caption{\label{fig:2} (a) Dependence $|P_0^{\prp}|^2$ on $|E_0^{\prp}|^2$ at $\Omega=-0.1$. Middle branch inside the bistability loop corresponds to the solutions unstable with respect to small perturbations. Arrows show the transitions of the system's state between the stable branches. (b) Profile of the function Re~$P_n^{\prp}$ for the stationary kink obtained by numerical simulations of Eq.~(\ref{dynamic}) at $\Omega=-0.1$ and $|E_0^{\prp}|^2=5.22\times10^{-5}$.}
\label{fig2}
\end{figure}

In general, there are two fundamental nonlinear phenomena inherent in our system: modulational instability and bistability. Here we discuss the letter phenomenon. When the spatially distributed system demonstrates bistable response, it can support simultaneously two different stable states forming switching waves or kinks. However, to observe kinks and study its properties, one should satisfy one more condition - the stability criterium, i.e. provide a range of the external field intensities (lying inside the bistability region) which is free from modulational instability. Our findings show that for an array of nanoparticles this condition is fulfilled for the transversal excitation only. Therefore, we, next, consider just this case.

To begin with, we excite the infinite array by an homogenous electric field with the transversal polarization: ${\bf E}_n=(E_0^{\prp},0)$. In this case, all particle dipole moments remain the same, $P_n^{\prp}=P_0^{\prp}$, and the stationary solution of Eq. (\ref{dynamic}) can be written as follows
\begin{equation}\label{initial state}
\left(-i\gamma+\Omega+\sum_{j=1}^\infty A_j^{\prp}+|P_0^{\prp}|^2 \right) P_0^{\prp}= E_0^{\prp},
\end{equation}
where $$A_j^{\prp}=\eta\left(-\frac{1}{j^{3}}-\frac{i k_0 d}{j^{2}} + \frac{(k_0 d)^2}{j} \right) \exp{(-ik_0 d j)}.$$ Transition from $G_{n,m}^{\prp}$ to $A_j^{\prp}$ has been made via the replacement $|n-m|=j$ and taking into account symmetry structure of the series.
When $$\Omega<-{\rm Re}\sum_{j=1}^\infty A_j^{\prp} - \sqrt{3}\left( \gamma - {\rm Im}\sum_{j=1}^\infty A_j^{\prp} \right),$$ the polarization $P_0^{\prp}$ becomes a three-valued function of $E_0^{\prp}$ leading to bistability (see Fig.~\ref{fig2}(a)).

\begin{figure}[t]
\centerline{\mbox{\resizebox{13cm}{!}{\includegraphics{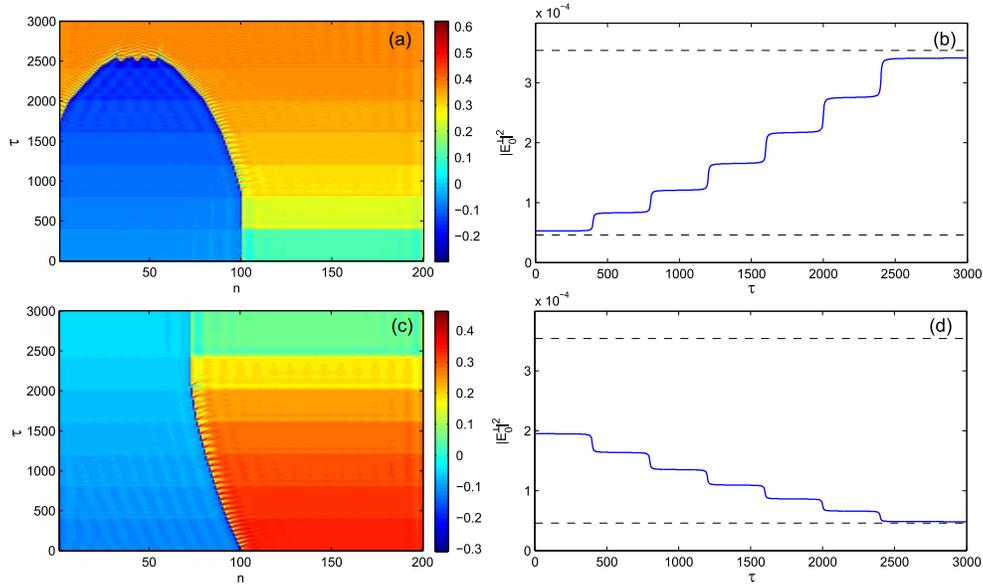}}}}
\caption{\label{fig:3} (a) (Media 1) and (c) (Media 2) demonstrate dynamics of Re~$P_n^{\prp}$ obtained numerically from Eq.~(\ref{dynamic}) at $\Omega=-0.1$. (b,d) External light intensities corresponding to the cases (a) and (c), respectively. Dashed lines mark the bistability region.}
\label{fig3}
\end{figure}
\begin{figure}[t]
\centerline{\mbox{\resizebox{8cm}{!}{\includegraphics{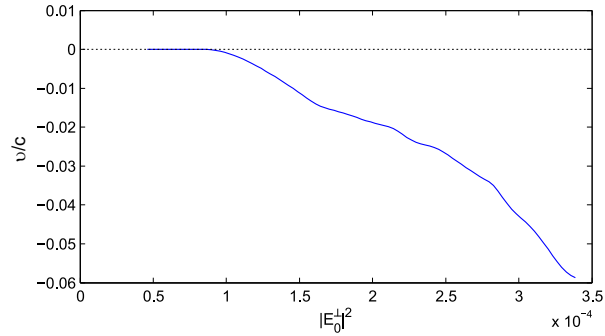}}}}
\caption{\label{fig:4}  Normalized kink velocity vs. the intensity of applied field $|E_0^{\prp}|^2$ at $\Omega=-0.1$.}
\label{fig4}
\end{figure}
To analyze switching waves, we perform numerical simulations of Eq. (\ref{dynamic}) for a finite array (with 200 nanoparticles) at initial
conditions corresponding to different branches of the homogeneous stationary solution (\ref{initial state}) over different parts of the array. In Fig.~\ref{fig2}(b) the profile of the stationary kink is presented. However, our calculations show that kinks can move as well. We have investigated impact of the field intensity on temporal kink dynamics. Characteristic results are
summarized in Figs.~\ref{fig3}(a,c). To simplify analysis, we fixed $\Omega$ and set $E_0^{\prp}$ as a step-like function of time, as shown in Figs.~\ref{fig3}(b,d). Constant $E_0^{\prp}$ results in the constant kink velocity, $\upsilon$. That is why a step-like time dependency of $E_0^{\prp}$ allows us to control $\upsilon$ overall bistability zone.

Figure~\ref{fig3}(a) demonstrates dynamical behavior of the system when $E_0^{\prp}$ increases from the lower to the upper threshold of the bistability region, as shown in Fig.~\ref{fig3}(b). In this case the kink state changes from standing to moving, and for $|E_0^{\prp}|^2$ lying close to the middle of the bistability region the second kink is appeared at the left edge of the array, moving towards the first one. This is the straight consequence of cooperation between edge effects and modulational instability, which, in the end, crushes both kinks.

To avoid appearance of modulational instability, we also carry out the simulation when $E_0^{\prp}$ decreases from the middle to the lower threshold of the bistability region. The result is presented in Fig.~\ref{fig3}(c). Here the kink velocity changes from the finite value to zero, and modulational instability growth is not observed.

Figure~\ref{fig4} shows the kink velocity as a function of the external field intensity $|E_0^{\prp}|^2$. We notice that there exists
a finite range of the field intensities where the kink does not move, i.e. $\upsilon=0$.  Generally speaking, this is a common property
of nonlinear discrete systems \cite{kivshar} where the effective periodic potential created by the lattice requires a finite value of the applied external force to start kink's motion.  The existence of the kinks with zero velocity suggest a straightforward way to construct spatially localized modes by taking two standing kink of the opposite polarity. The resulting localized modes form the continuous families, and they resemble
the similar localized modes discussed earlier for metal-dielectric multilayers \cite{prl_zhang,skryabin_oe_2011,skryabin_ol_2010} and arrays of metal nanowires~\cite{prl_panoiu}. On the other hand, we may expect generating such subwavelength plasmon solitons by considering interaction of two slowly moving kinks of the opposite polarity.

\begin{figure}
\centerline{\mbox{\resizebox{13cm}{!}{\includegraphics{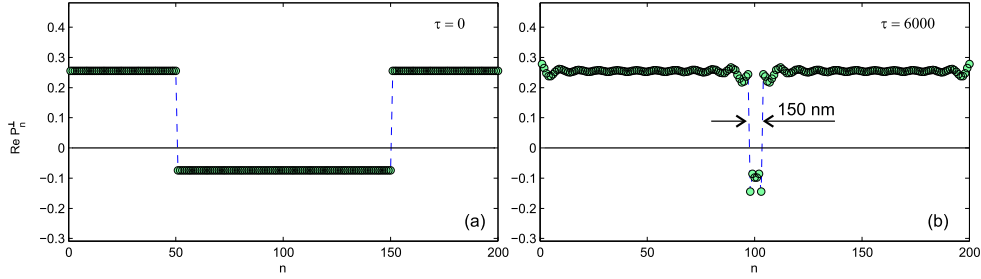}}}}
\caption{\label{fig:5} Generation of a plasmon soliton via interaction of two kinks with the opposite polarities.
(a) Initial condition used for numerical simulations of Eq.~(\ref{dynamic}) at $|E_0^{\prp}|^2=9.27\times10^{-5}$ and $\Omega=-0.1$.
(b) (Media 3) The frozen frame of Re~$P_n^{\prp}$ showing the profile of the subwavelength dissipative plasmon soliton formed by two interacting kinks.}
\label{fig5}
\end{figure}

Here, we demonstrate this latter approach. We study numerically the interaction of two kinks of the opposite polarities, and select the value of the external field to realize the kinks with the near-zero velocity $\upsilon$ [see  Fig.~\ref{fig5}(a)]. We find that the kinks start moving slowly towards each other, then stop and form a stable dissipative plasmon soliton, as shown in Fig.~\ref{fig5}(b). The generated plasmonic mode is localized  only on 6 particles, and its width is about 150 nm. Because $\Omega=-0.1$ corresponds to the radiation wavelength $\lambda=440$~nm, this plasmon soliton is truly subwavelength because it extends for $0.34\lambda$.

It remains open if the kink's motion may demonstrate bistability, as was recently revealed for  a  similar nonlinear system  with local coupling [23]. This will be a subject for future studies.
\section{Conclusions}
We have demonstrated that an array of interacting nonlinear metal nanoparticles can support a novel type of nonlinear localized modes,
plasmonic kinks, which describe switching waves connecting two different states of polarization of the metal particles. We have
generated various types of such kinks, analyzed numerically their velocity as a function of the external field intensity, and revealed that the array of metal nanoparticles can support deeply subwavelength dissipative plasmon solitons which in our case can be regarded
as a bound state of two kinks of the opposite polarity.

\section*{Acknowledgement}
The authors acknowledge a support from Australian Research Council and Russian Ministry of Education and Science, as well as fruitful discussions with A.A.~Zharov and N.N.~Rosanov.

\end{document}